\documentclass[twocolumn,showpacs,preprintnumbers,amsmath,amssymb,superscriptaddress]{revtex4}
\pdfoutput=1

\usepackage{graphicx}
\usepackage{dcolumn}
\usepackage{bm}

\def\be{\begin{equation}}
\def\ee{\end{equation}}
\def\bea{\begin{eqnarray}}
\def\eea{\end{eqnarray}}

\newcommand{\lsim}{ \mathop{}_{\textstyle \sim}^{\textstyle <} }

\newcommand{\kms}{{\rm km/s}}
\newcommand{\kev}{{\rm keV}}

\newcommand{\gev}{{\rm GeV}}

\begin{document}

\preprint{MCTP-08-58}

\title{Using the Energy Spectrum at DAMA/LIBRA to Probe Light Dark Matter}

\author{Spencer Chang}%
\affiliation{Center for Cosmology and Particle Physics, New York University, New York, NY 10003}
\author{Aaron Pierce}
\affiliation{Michigan Center for Theoretical Physics,  University of Michigan, Ann Arbor MI, 48109}
\author{Neal Weiner}%
\affiliation{Center for Cosmology and Particle Physics, New York University, New York, NY 10003}

\date{\today}

\begin{abstract}
A weakly interacting massive particle (WIMP) weighing only a few GeV has been invoked as an explanation for the signal from the DAMA/LIBRA experiment.   We show that the data from DAMA/LIBRA are now powerful enough to strongly constrain the properties of any putative WIMP.  Accounting for the detailed recoil spectrum, a light WIMP with a Maxwellian velocity distribution and a spin-independent (SI) interaction cannot account for the data. Even neglecting the spectrum, much of the parameter space is excluded by limits from the DAMA unmodulated signal at low energies. Significant modifications to the astrophysics or particle physics can open light mass windows.

\end{abstract}

\maketitle

\section{\label{sec:Intro} Introduction}

The DAMA/LIBRA NaI(Tl) scintillation experiment \cite{Bernabei:2008yi} has used the annual modulation technique \cite{Drukier:1986tm,Freese:1987wu} to search for dark matter (DM).  They now find a modulation of over 8$\sigma$ with a period and phase consistent with a DM signal.  
However, in some models of DM, it is not trivial to square this positive result with the null results from other direct detection experiments.  

Recent investigations \cite{Foot:2008nw,Feng:2008dz,Petriello:2008jj,Bottino:2008mf}, updating the discussion of \cite{Gondolo:2005hh} note that light DM, with mass of a few GeV  might reconcile the DAMA/LIBRA data with constraints from other experiments, e.g., \cite{CDMS,XENON}.   In addition, a DM candidate with mass $\sim$ GeV is tantalizing -- it might give insight into the ratio of the DM density to the density of baryons.


In this note, we point out that the statistics of the DAMA/LIBRA data are now sufficiently powerful that an explanation of the DAMA/LIBRA data must now go beyond just fitting the overall modulation rate.  Additional  self-consistency checks on the light WIMP scenario are now possible.  We discuss two such checks. 

First,  DAMA/LIBRA has now measured the modulation rate as a function of the observed recoil energy.  This spectrum contains valuable information.  
Simple kinematics indicate that the spectra of nuclei recoiling against a WIMP are sensitive to the mass of the WIMP.   Thus, fitting the observed energy spectrum constrains the mass of the candidate WIMP particle.   Another constraint can be derived by looking at the total (unmodulated) rate of observed events with low energy recoils.  Some WIMP candidates will provide more events than the  {\it total} number of observed events at low energies, despite a presumably sizable background.  

These two constraints are powerful probes of the light WIMP region, and effectively exclude this interpretation if a Maxwellian velocity distribution with standard parameters is assumed.  Modifications to this assumption, in particular DM streams, can open up small regions of allowed parameter space.

\section{\label{sec:Rates} Modulation Rates and DAMA}
To calculate the detection rates, we follow the standard formalism reviewed in \cite{JKG,LewinSmith}.  The differential scattering rate is given by:
\be
\frac{dR}{dE_R} = \frac{M_{N} N_{T} \rho_{\chi} \sigma_n}{2 m_\chi \mu_{ne}^2} F^{2}  \frac{(f_{p} Z + f_n (A-Z))^{2}} {f_n^2} \int_{v_{min}} \frac{f(v)}{v}.
\ee
Here, $M_{N}$ is the mass of the target nucleus, $N_{T}$ is the number of target nuclei in a detector, and $\mu_{ne}$ is the reduced mass of the WIMP-nucleon system.  
We take the local DM density as $\rho_{\chi} =0.3$ GeV/cm$^{3}$.  $F^{2}$ is a nuclear form factor, while $\sigma_n$ represents the cross section to scatter on a nucleon at zero momentum transfer.  The relative coupling to protons and neutrons are given by $f_{p}$ and $f_{n}$.  We take these couplings equal.
This choice does not have a large effect.  We use the Helm form factor \cite{LewinSmith,JKG}.   Because the $F^{2} \approx 1$ for the recoils in the low mass window, we do not expect our results to be sensitive to the detailed form of the form factor.

The modulation signal arises from the seasonal differences in the speed distribution, $f(v)$. When the earth moves with the sun through the WIMP halo, the scattering rate above a given threshold is higher: the flux of WIMPs increases and the higher relative velocities favor harder scatters.  We take the WIMP halo to have a Maxwell-Boltzmann (MB) distribution with dispersion $v_{0} = 220$ km/s, cut-off by an escape velocity of  600 km/s in the frame where the halo is isotropic.  This value of the escape velocity is consistent with recent observations from the RAVE survey \cite{RAVE}, which quotes a value of $498\; \rm{km/s} < v_{esc} <608\; \rm{km/s}$ \footnote{Our conclusions are not affected by variations of these parameters within the allowed range.}.  The earth moves through this halo at approximately 240 km/s, so the highest velocity WIMPs in this boosted frame have velocities of approximately 840 km/s.  In doing our numerical work, we take the earth velocity from \cite{LewinSmith} and the solar velocity relative to the rotation velocity (taken to be $v_{0}$) from \cite{SolarV}.

\begin{figure*}
\includegraphics[width=0.45\textwidth]{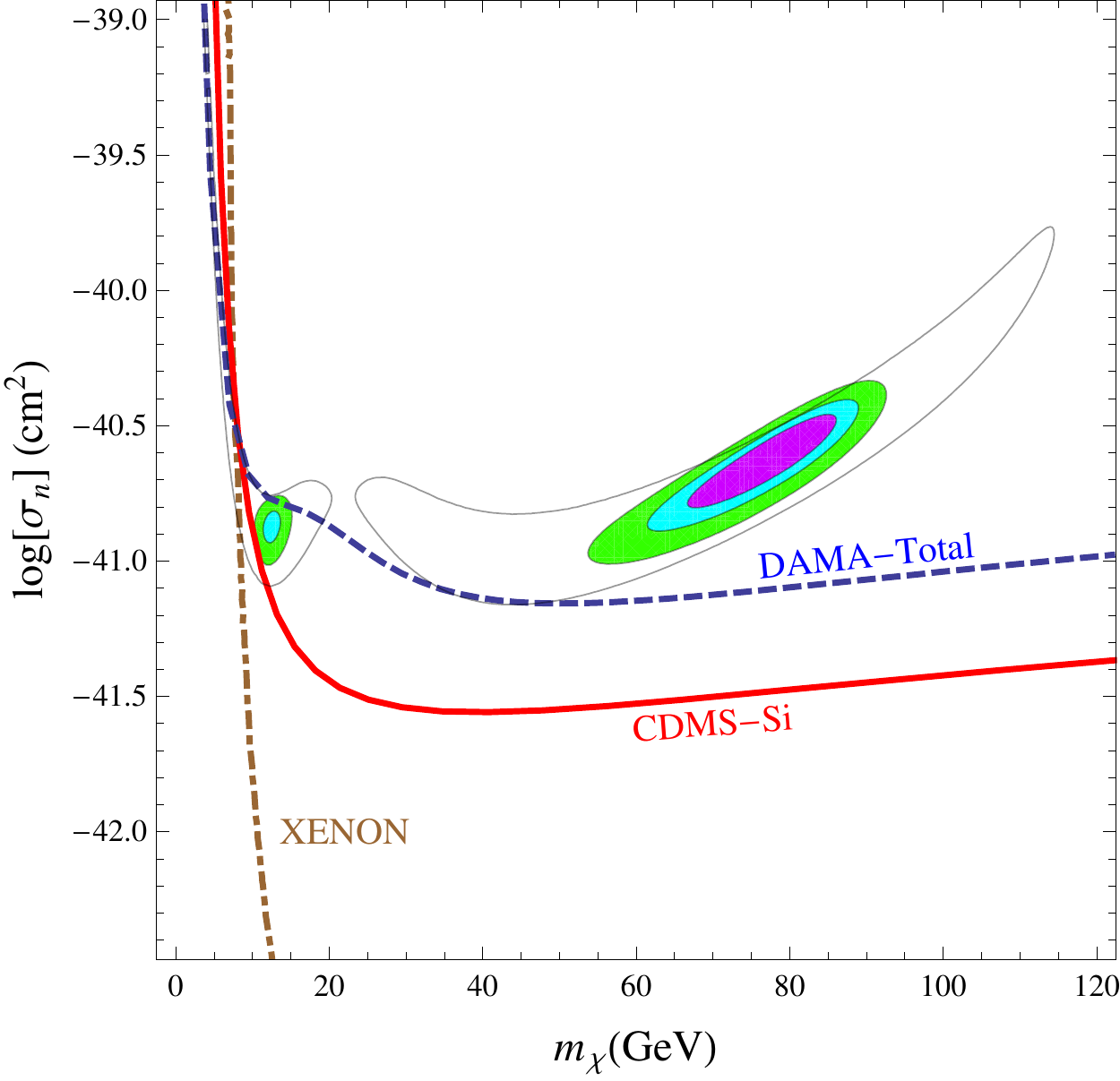}
\includegraphics[width=0.45\textwidth]{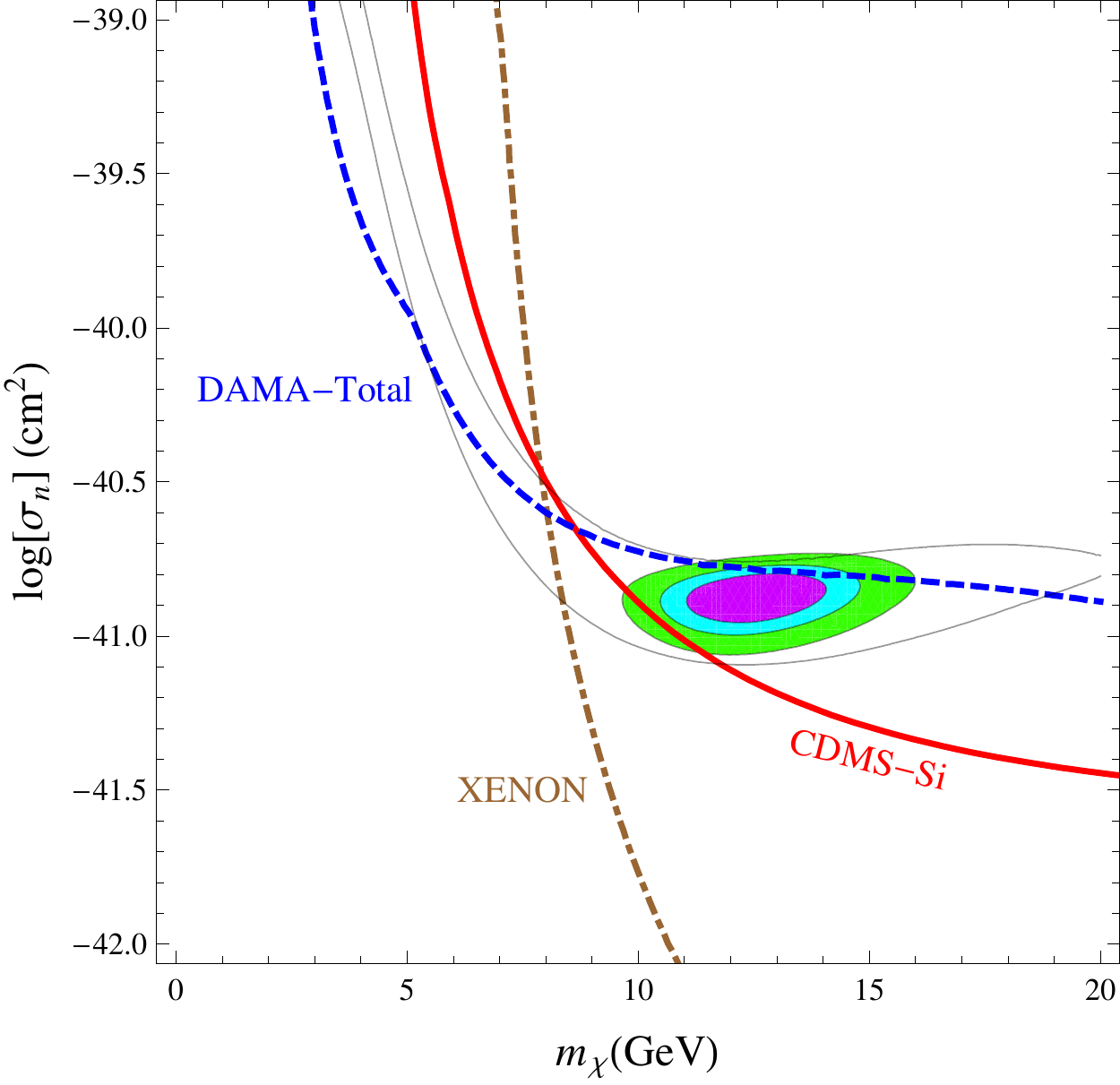}
\caption{We show the region consistent with the DAMA/LIBRA modulation signal  (at 68\%, 90\% and 99\% CL),  accounting for  the detailed modulation energy spectrum between 2-6 keVee and the single bin modulation between 6-14 keVee (colored contours). Around this, we show the envelope (gray line) consistent with the DAMA/LIBRA modulation signal at 99\% CL, fitting the total modulation rate from the two bins, 2-6 keVee and 6-14 keVee (i.e., no spectral information).   Also shown are bounds calculated by considering data from the CDMS-II (Si) and XENON10 experiments.  The limit curve labeled ``DAMA-Total'' arises from the self-consistency with the total (unmodulated) rates in the DAMA-LIBRA data, as described in Section~\ref{sec:DAMAvDAMA}. At right, we zoom in on the light mass region. When zooming in, we study variations about the minimum $\chi^2$ in the low mass region, as described in the text.}
\label{fig:NewContours}
\end{figure*}

DAMA/LIBRA presents the modulation signal as a function of the observed energy, measured in keV electron equivalent (keVee).  Conversion between true nuclear recoil energy (keV) and observed energy (keVee) requires a quenching factor ($QF_{Na}$ =0.3, $QF_{I}$= 0.09).   We have included Gaussian energy smearing using the value: $ \sigma(E)/E =0.448/\sqrt{E} + 9.1 \times 10^{-3}$ (all energies in keVee) \cite{DAMAApparatus}.   It has been noted \cite{Petriello:2008jj} that  a low mass window consistent with other experiments, relies on the  ``channeling'' effect \cite{Bernabei:2007hw}.   When channeling occurs a recoiling nucleus interacts with the crystal only electromagnetically, efficiently transferring its energy to photoelectrons. 
The result is that  the full energy of a nuclear recoil is observed as scintillation light, rather than the quenched value.  We include channeling following the parameterization of \cite{Foot:2008nw}, also used in \cite{Petriello:2008jj}.
In this approximation, if channeling occurs in a given event, then all recoil energy is observed without quenching \footnote{We have checked that if this approximation were relaxed (perhaps only a larger, but still non-unary fraction of the energy were observed), then, while quantitative differences exist, our basic conclusions are unchanged.}.

Fitting the DAMA modulation between 2-6 keVee and 6-14 keVee as two bins, as done in \cite{Gondolo:2005hh,Petriello:2008jj},  we find a large region consistent at 99\% CL,
(gray unshaded contour in Fig.~\ref{fig:NewContours}).  There are two qualitatively distinct regions.  One is  centered roughly around 80 GeV, where the scattering dominantly proceeds off of the iodine nuclei.  This region is clearly excluded by other direct detection experiments.  The second is at lower mass, extending down to a few GeV, where the scattering is dominantly off sodium nuclei.   This DM is too light to cause recoils on iodine above detection threshold.  Because a typical SI detection cross section scales with the square of atomic number, the cross section per nucleon for a WIMP that scatters off of sodium must be large.   It is this low mass, high cross-section region that appeared consistent with all known direct detection experiments \footnote{Note that our region differs somewhat from the contour reported in \cite{Petriello:2008jj}, we believe this is due to the way the escape velocity cutoff was implemented there.}.    We now show that detailed spectral information modifies this region so that it is disfavored by current experimental results.

\section{\label{sec:DAMAvDAMA} DAMA vs. DAMA}

\begin{figure}[t]
\includegraphics[width=0.45\textwidth]{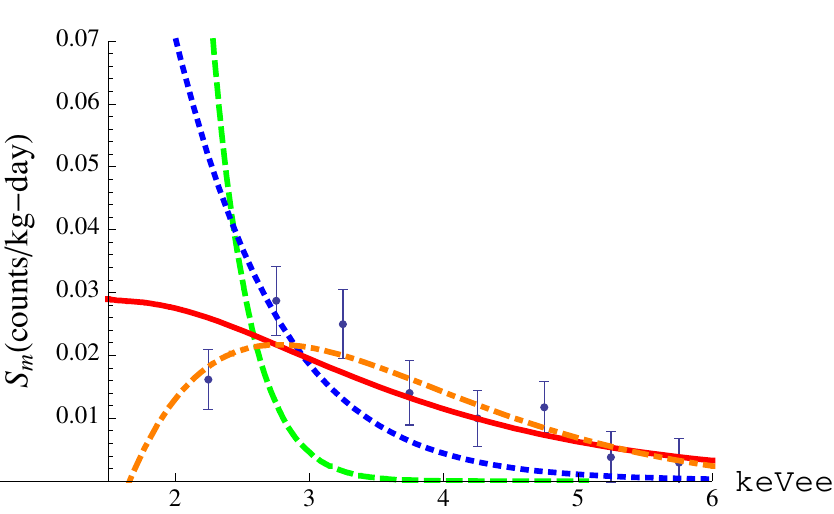}
\caption{We show the modulation spectra for the best fit point where scattering off iodine dominates, $m_{\chi}$ = 77 GeV (dot-dashed orange), and three points where scattering off of sodium dominates.  The best fit point off sodium is $m_{\chi}$ = 12 GeV (solid red).    We also show $m_{\chi} = 2$ GeV (dashed green) and $m_{\chi}=7$ GeV (dotted blue).  The points with error bars are the published DAMA/LIBRA data.} 
\label{fig:SpectrumMass}
\end{figure}

In Fig.~\ref{fig:SpectrumMass}, we show the observed modulation signal $S_{m}$ vs. observed energy for a WIMP with SI interaction at DAMA/LIBRA.  Two of the plotted modulation spectra are for the masses that provide the best fits when the WIMP scatters dominantly on iodine ($m_{\chi}$  = 77 GeV), and on sodium ($m_{\chi}$ = 12 GeV).  Our fit is to a nine bin $\chi^2$ from the eight .5 keVee bins between 2-6 keVee, where modulation is seen, and one bin from 6-14 keVee, where it is consistent with no modulation.  The fit in the iodine scattering region is excellent ($\chi^{2}_{min}$ = 4.21 for 7 dof, $p=0.76$).  The fit in the sodium region is worse: $(\chi^{2}_{min}$ = 7.54, $p=0.37$).   From Fig.~\ref{fig:SpectrumMass}, it is easy to see why this is so.  At the best fit point in the low mass window, the DM is too light to have most of its recoils in the region of interest (2-6 keVee). The amount of modulation is still rising as one goes to lower energies --- in conflict with the DAMA/LIBRA data.  While one might think that this could be remedied by moving to a higher mass, say 20 GeV, this approach does not succeed.  As the mass moves above 12 GeV, a contribution coming from iodine scattering begins to move into the low end of the observed energy region, spoiling the fit. Also plotted in Fig.~\ref{fig:SpectrumMass} are spectra for WIMP masses of 2 and 7 GeV for comparison.  

The 68\%, 90\%, and 99\% CL ($\Delta \chi^2 <$ 2.3, 4.61, 9.21) contours consistent with our nine bin DAMA/LIBRA $\chi^2$ function are shown in Fig.~\ref{fig:NewContours}.  Both regions shrink dramatically compared to the two bin $\chi^2$.  In the left panel, the $\Delta \chi^2$ is with respect to the global best fit point at 77 GeV.  In the right panel, we concentrate on the low mass region and have defined $\Delta \chi^2$ relative to the low mass best fit point of 12 GeV.  Note this region is confined to masses above 10 GeV.  This can be understood from examining the recoil spectra for the light WIMPs in  Fig.~\ref{fig:SpectrumMass}.  For a fixed overall modulation rate, sub-10 GeV WIMPs predict too little modulation above a couple of keVee: they simply do not have enough mass to cause recoils of this size.  

In Fig.~\ref{fig:NewContours}, we have superimposed 90\% limit contours from both the CDMS \cite{CDMS} and XENON \cite{XENON} experiments.   We only show the CDMS contour  relevant for low masses, corresponding to  data taken with the silicon detectors in the Soudan mine, which have a  7 keV threshold.  We have recalculated limits using the astrophysical parameters described here.  For the XENON experiment, we account for the energy dependent efficiencies as described in \cite{XENON} and to set the limit, apply the maximum gap method \cite{Yellin:2002xd} to the energy recoil range of 4.5-26.9 keV. We see that the light SI DAMA/LIBRA mass region is excluded once the modulation spectrum is taken into account.

Another constraint can be imposed by looking at the total (unmodulated) counts at low energies at DAMA/LIBRA.  For a WIMP, the number of recoil events increases with decreasing energy.  Since DAMA/LIBRA cannot distinguish background events from signal events in this sample, it is clear that the predicted number of WIMP events should not excessively exceed their total number of counts in any bin.

We require that the unmodulated rate in each bin from $0.75-4$ keVee not exceed the observed values within their 90\% error \cite{Bernabei:2008yi}. We show this constraint in Fig.~\ref{fig:NewContours}, labeled  ``DAMA-Total.''  The allowed region lies below this curve.  This constraint does not greatly impinge upon the allowed region from our nine bin $\chi^2$ that accounts for the spectral details.   Its constraint is most striking if one considers the allowed region of the two bin fit (see Fig.~\ref{fig:NewContours}).

Since the modulation in the low $(2-2.5\; \kev)$ bin seems to be the most constraining, one can consider whether it is overly biasing our analysis. For instance, one could worry if DAMA/LIBRA were to restate the efficiency in the lowest bin, this might completely change our results. We have explored such effects by various methods: tripling the error bar on the lowest bin, merging the entire range into a $2-3\; \kev$ bin, and discarding the lowest bin entirely. We find that only the last option (discarding the lowest bin) opens up a region of parameter space, with a point allowed with $\chi^2 =9.14$ for 6 dof ($p=0.17$). This point also has a unmodulated rate that is close to saturating the observed rate.

\section{Variations from Astrophysics and Particle Physics}
Thus far, we assumed a MB halo and an elastic, SI interaction. 
Relaxing these assumptions could enlarge the region at light masses, so that modulation arises with an appropriate spectrum, consistent with other experiments. 

Let us begin by considering astrophysical modifications to the velocity distribution. Kinematics informs us of what modifications are needed. To scatter with nuclear recoil energy $E_R$, a WIMP must have a minimum velocity $\beta_{min} = \sqrt{M_N E_R/2 \mu^2}$, where $\mu$ is the reduced mass of the WIMP-nucleus (not nucleon) system.  Consider the channeled possibility, for which the velocity requirements are weakest: for scattering on sodium, with recoil energy of 4.5 keV (the highest bin with significant modulation), one finds $\beta_{min} c \approx 1140, 790, 620 \;\kms$ for $m_\chi = 2,3,4 \;\gev$.
If halo particle velocities approximately follow a MB distribution, the most significant deviations naturally occur for the highest velocities, where recent infall and streams may not have fully virialized. As such, the lightest particles are the most likely to have allowed regions opened by such deviations from a MB distribution.

\begin{figure}
\includegraphics[width=0.45\textwidth]{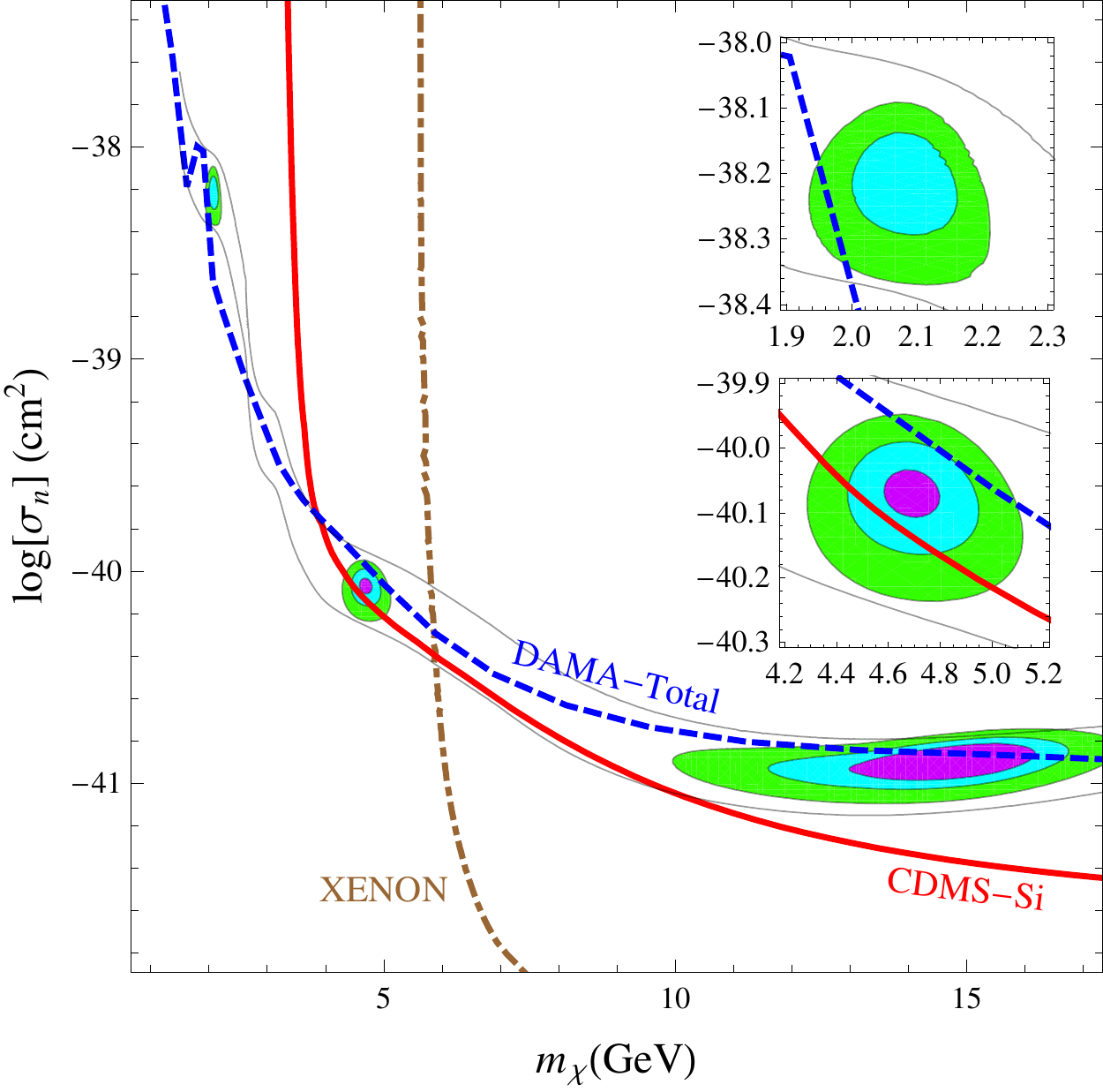}
\caption{Similar to Fig.~\ref{fig:NewContours}, but incorporating a stream of DM, with $ v_{str} = 900 \; \kms,$ and $\sigma_{str}=20 \; \kms$.  Insets:  magnification of parameter space near 2 GeV and 4 GeV.  The 2 GeV relies on both channeling and the stream, the 4 GeV region arises from unchanneled events from the stream.}
\label{fig:streampoint}
\end{figure}

 One modification to the halo is to include streams \cite{Gelmini:2000dm,Stream}.   We investigated a wide range of streams, varying its velocity $-1200 \; \kms < v_{str} < 1200 \;\kms$ and dispersion $10 \;\kms < \sigma_{str} < 50 \;\kms$.  Our stream is such that for positive (negative) $v_{str}$, the stream is directly against (with) the Earth's velocity as given in the sun's rest frame.    
 Because we limit ourselves to small perturbations on MB (we take the stream to have 3\% of the density of the MB contribution), only streams well away from the bulk of the MB distribution make a difference.   Shown in Fig.~\ref{fig:streampoint}, for $v_{str}= 900 \; \kms,\, \sigma_{str}=20 \;\kms$, a channeled region near $m_\chi=2$ GeV appears with a marginal fit of $\chi^2 = 10.3$ and $p= 0.17$, as well as an unchanneled region near $m_\chi=4$ GeV with a fit of $\chi^2=9.77$ and $p= 0.2$.  These low mass regions are allowed only for streams that are not bound to the galaxy, with $v_{str} \gtrsim 800 \;\kms$.  As such, we view these modifications as significant from the typical MB distribution.  Similar light regions for negative $v_{str}$  are ruled out by the DAMA unmodulated limit.  

Direct detection experiments must go to very low thresholds to probe these theories.  Comparisons within the context of MB halos will not be informative \footnote{This is not to say that limits should not be presented with MB halos.  Rather, as we know that MB gives a poor fit to the DAMA data at low masses, comparisons outside of MB are essential.}. Rather, experiments should focus on probing a comparable range of velocity space. 
Probing the DAMA/LIBRA region of parameter space for nuclear recoils, requires $E_{thresh} \simeq 4 \; {\rm keV} \times \left( \frac{M_{Na}}{M_{N}} \right)$, where $M_N$ is the target nucleus in question.
For Si, a 3.3 keV threshold would be sufficient, while for Ge, and Xe, 1.25 keV  and 0.7 keV thresholds are required, respectively.  Low thresholds for both Si and Ge are being explored \cite{mahapatra} and given projections from CoGeNT \cite{Aalseth:2008rx}, a recent low threshold germanium experiment, this low mass region could be probed even in the presence of non-MB distributions.

 A simple modification of the particle physics would be the inelastic DM scenario (iDM) \cite{IDM}, wherein scattering occurs via a transition to a slightly heavier state, $m_{\chi^*}-m_\chi \equiv \delta \sim \beta^2 m_\chi$. The WIMP must now be moving fast enough to excite to the heavier state, i.e., 
\begin{eqnarray}
\beta_{\rm min} &=& \sqrt{\frac{1}{2 m_N E_R}} \left( \frac{m_N E_R}{\mu} + \delta \right).
\label{eq:inelasticbetamin}
\end{eqnarray}
Such an inelastic WIMP at heavier masses ($\sim$ 100 GeV) has recently been shown to be consistent with DAMA's signal and the limits from other experiments \cite{IDMNew}.

\begin{figure}[t]
\includegraphics[width=0.45\textwidth]{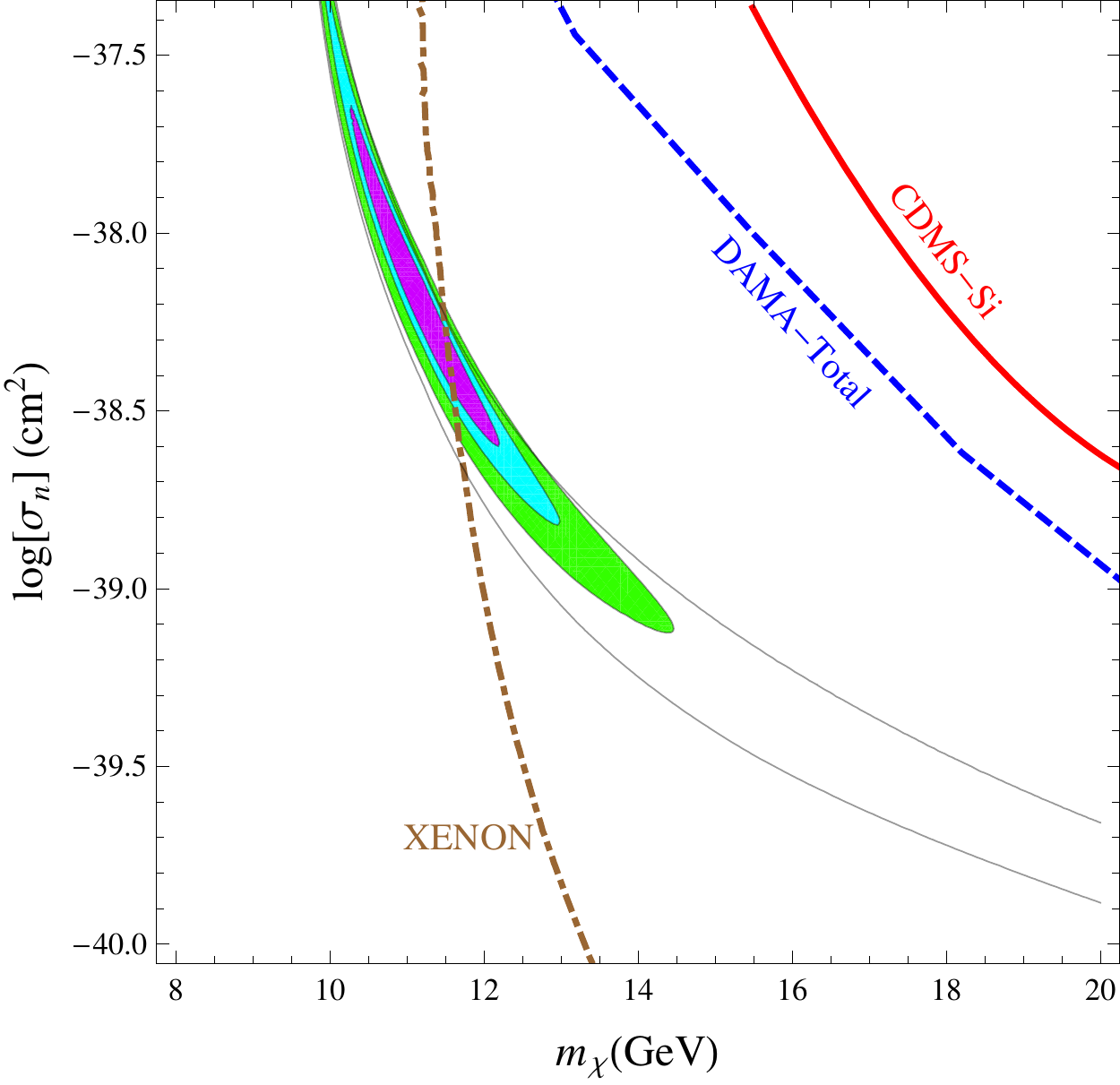}
\caption{Similar to Fig.~\ref{fig:NewContours}, but including only inelastic scatterings, with $\delta = 35 \;\kev$.}
\label{fig:inelastic}
\end{figure}

Such scatterings have a recoil spectrum which can peak at 2-4 keVee, while being zero at low energies \cite{IDM}. We have explored inelastic scenarios, and find regions with $\delta \sim 30 \; \kev$, $m_\chi \sim 10 \; \gev$ and $\sigma_n \sim 10^{-39} \;{\rm cm^2}$ that are consistent with other experiments. XENON10 is the most constraining. We show the allowed region for $\delta = 35 \; \kev$ in Fig. \ref{fig:inelastic}.

Building an inelastic model with such a large $\sigma_{n}$ is difficult.  In the prototypical inelastic model, scattering arises from $Z$ exchange.  A light, vector-like Dirac neutrino would scatter off sodium with $\sigma_n \approx 1.5 \times 10^{-39}\; {\rm cm^2}$.  But since the number of neutrinos is limited by LEP to be $n_\nu = 2.9840 \pm0.0082$ \cite{pdg}, the $Z$-DM coupling must be suppressed, for instance by mixing with a singlet state. This bounds the cross section  $\sigma_n \lsim 1.3 \times 10^{-41}\; {\rm cm^2}$. Thus, there is significant tension between a viable model of light inelastic DM, and the overall signal size.

Nonetheless, modifications to the astrophysics and particle physics can re-open some windows in the light WIMP regime, with astrophysics likely opening $m_\chi \sim 2-5 \; \gev$ regions, and inelasticity opening $m_\chi \sim 10\; \gev$ windows. Both of these warrant further investigation.

\section{\label{sec:Conclusion} Conclusions}
The recent update from the DAMA/LIBRA collaboration has added significance to the modulated signature. Detailed spectral data should now be considered for any model proposed to explain the signal. As it appears that modulation extends up to at least 5 keVee, the spectrum places very strong constraints on the properties of a standard WIMP in a MB velocity distribution. Additionally, because the signal from a standard WIMP rises very rapidly at low recoil energies, one must insure that the {\em unmodulated} signal predicted does not exceed the observed levels at DAMA/LIBRA.  

Fitting to the spectral data, one finds two regions for a standard WIMP in a Maxwellian halo which explain the spectrum. The high mass region conflicts with other direct detection experiments and the unmodulated low-energy rate at DAMA/LIBRA. The light mass region gives a poorer fit to the spectrum, but is still acceptable 
and is largely consistent with limits from the DAMA unmodulated signal. However,  90\% CL limits from CDMS-Si and XENON-10 exclude this region.  

Modifications to astrophysics or particle physics can open light mass windows. Streams can open regions in the $m_\chi \sim 2 -5\; \gev$ range, although for very particular choices of parameters. Inelasticity can broaden areas at higher masses, but faces model-building challenges.   Future direct detection experiments should push to explore both of these regions.

\begin{acknowledgments}
We would like to thank Frank Petriello, Chris Savage and Kathryn Zurek for useful discussions and comments.  NW and AP acknowledge the hospitality of the Kavli Institute of Theoretical physics, where this work was begun and the the Aspen Center for Physics, where much of this work was completed.  AP is supported by NSF CAREER Grant NSF-PHY-0743315.  NW and SC are supported by  NSF CAREER grant PHY-0449818 and DOE grant  DE-FG02-06ER41417. 
\end{acknowledgments}

\bibliography{LDM}

\begin{thebibliography}{24}
\expandafter\ifx\csname natexlab\endcsname\relax\def\natexlab#1{#1}\fi
\expandafter\ifx\csname bibnamefont\endcsname\relax
  \def\bibnamefont#1{#1}\fi
\expandafter\ifx\csname bibfnamefont\endcsname\relax
  \def\bibfnamefont#1{#1}\fi
\expandafter\ifx\csname citenamefont\endcsname\relax
  \def\citenamefont#1{#1}\fi
\expandafter\ifx\csname url\endcsname\relax
  \def\url#1{\texttt{#1}}\fi
\expandafter\ifx\csname urlprefix\endcsname\relax\def\urlprefix{URL }\fi
\providecommand{\bibinfo}[2]{#2}
\providecommand{\eprint}[2][]{\url{#2}}

\bibitem[{\citenamefont{Bernabei et~al.}(2008{\natexlab{a}})}]{Bernabei:2008yi}
\bibinfo{author}{\bibfnamefont{R.}~\bibnamefont{Bernabei}} \bibnamefont{et~al.}
  (\bibinfo{collaboration}{DAMA}) (\bibinfo{year}{2008}{\natexlab{a}}),
  \eprint{0804.2741}.

\bibitem[{\citenamefont{Drukier et~al.}(1986)\citenamefont{Drukier, Freese, and
  Spergel}}]{Drukier:1986tm}
\bibinfo{author}{\bibfnamefont{A.~K.} \bibnamefont{Drukier}},
  \bibinfo{author}{\bibfnamefont{K.}~\bibnamefont{Freese}}, \bibnamefont{and}
  \bibinfo{author}{\bibfnamefont{D.~N.} \bibnamefont{Spergel}},
  \bibinfo{journal}{Phys. Rev.} \textbf{\bibinfo{volume}{D33}},
  \bibinfo{pages}{3495} (\bibinfo{year}{1986}).

\bibitem[{\citenamefont{Freese et~al.}(1988)\citenamefont{Freese, Frieman, and
  Gould}}]{Freese:1987wu}
\bibinfo{author}{\bibfnamefont{K.}~\bibnamefont{Freese}},
  \bibinfo{author}{\bibfnamefont{J.~A.} \bibnamefont{Frieman}},
  \bibnamefont{and} \bibinfo{author}{\bibfnamefont{A.}~\bibnamefont{Gould}},
  \bibinfo{journal}{Phys. Rev.} \textbf{\bibinfo{volume}{D37}},
  \bibinfo{pages}{3388} (\bibinfo{year}{1988}).

\bibitem[{\citenamefont{Feng et~al.}(2008)\citenamefont{Feng, Kumar, and
  Strigari}}]{Feng:2008dz}
\bibinfo{author}{\bibfnamefont{J.~L.} \bibnamefont{Feng}},
  \bibinfo{author}{\bibfnamefont{J.}~\bibnamefont{Kumar}}, \bibnamefont{and}
  \bibinfo{author}{\bibfnamefont{L.~E.} \bibnamefont{Strigari}}
  (\bibinfo{year}{2008}), \eprint{0806.3746}.

\bibitem[{\citenamefont{Petriello and Zurek}(2008)}]{Petriello:2008jj}
\bibinfo{author}{\bibfnamefont{F.}~\bibnamefont{Petriello}} \bibnamefont{and}
  \bibinfo{author}{\bibfnamefont{K.~M.} \bibnamefont{Zurek}}
  (\bibinfo{year}{2008}), \eprint{0806.3989}.

\bibitem[{\citenamefont{Bottino et~al.}(2008)\citenamefont{Bottino, Donato,
  Fornengo, and Scopel}}]{Bottino:2008mf}
\bibinfo{author}{\bibfnamefont{A.}~\bibnamefont{Bottino}},
  \bibinfo{author}{\bibfnamefont{F.}~\bibnamefont{Donato}},
  \bibinfo{author}{\bibfnamefont{N.}~\bibnamefont{Fornengo}}, \bibnamefont{and}
  \bibinfo{author}{\bibfnamefont{S.}~\bibnamefont{Scopel}}
  (\bibinfo{year}{2008}), \eprint{0806.4099}.

\bibitem[{\citenamefont{Foot}(2008)}]{Foot:2008nw}
\bibinfo{author}{\bibfnamefont{R.}~\bibnamefont{Foot}} (\bibinfo{year}{2008}),
  \eprint{0804.4518}.

\bibitem[{\citenamefont{Gondolo and Gelmini}(2005)}]{Gondolo:2005hh}
\bibinfo{author}{\bibfnamefont{P.}~\bibnamefont{Gondolo}} \bibnamefont{and}
  \bibinfo{author}{\bibfnamefont{G.}~\bibnamefont{Gelmini}},
  \bibinfo{journal}{Phys. Rev.} \textbf{\bibinfo{volume}{D71}},
  \bibinfo{pages}{123520} (\bibinfo{year}{2005}), \eprint{hep-ph/0504010}.

\bibitem[{\citenamefont{Akerib et~al.}(2006)}]{CDMS}
\bibinfo{author}{\bibfnamefont{D.~S.} \bibnamefont{Akerib}}
  \bibnamefont{et~al.} (\bibinfo{collaboration}{CDMS}), \bibinfo{journal}{Phys.
  Rev. Lett.} \textbf{\bibinfo{volume}{96}}, \bibinfo{pages}{011302}
  (\bibinfo{year}{2006}), \eprint{astro-ph/0509259}.

\bibitem[{\citenamefont{Angle et~al.}(2008)}]{XENON}
\bibinfo{author}{\bibfnamefont{J.}~\bibnamefont{Angle}} \bibnamefont{et~al.}
  (\bibinfo{collaboration}{XENON}), \bibinfo{journal}{Phys. Rev. Lett.}
  \textbf{\bibinfo{volume}{100}}, \bibinfo{pages}{021303}
  (\bibinfo{year}{2008}), \eprint{0706.0039}.

\bibitem[{\citenamefont{Jungman et~al.}(1996)\citenamefont{Jungman,
  Kamionkowski, and Griest}}]{JKG}
\bibinfo{author}{\bibfnamefont{G.}~\bibnamefont{Jungman}},
  \bibinfo{author}{\bibfnamefont{M.}~\bibnamefont{Kamionkowski}},
  \bibnamefont{and} \bibinfo{author}{\bibfnamefont{K.}~\bibnamefont{Griest}},
  \bibinfo{journal}{Phys. Rept.} \textbf{\bibinfo{volume}{267}},
  \bibinfo{pages}{195} (\bibinfo{year}{1996}), \eprint{hep-ph/9506380}.

\bibitem[{\citenamefont{Lewin and Smith}(1996)}]{LewinSmith}
\bibinfo{author}{\bibfnamefont{J.~D.} \bibnamefont{Lewin}} \bibnamefont{and}
  \bibinfo{author}{\bibfnamefont{P.~F.} \bibnamefont{Smith}},
  \bibinfo{journal}{Astropart. Phys.} \textbf{\bibinfo{volume}{6}},
  \bibinfo{pages}{87} (\bibinfo{year}{1996}).

\bibitem[{\citenamefont{Smith et~al.}(2007)}]{RAVE}
\bibinfo{author}{\bibfnamefont{M.~C.} \bibnamefont{Smith}}
  \bibnamefont{et~al.}, \bibinfo{journal}{Mon. Not. Roy. Astron. Soc.}
  \textbf{\bibinfo{volume}{379}}, \bibinfo{pages}{755} (\bibinfo{year}{2007}),
  \eprint{astro-ph/0611671}.

\bibitem[{\citenamefont{Dehnen and Binney}(1998)}]{SolarV}
\bibinfo{author}{\bibfnamefont{W.}~\bibnamefont{Dehnen}} \bibnamefont{and}
  \bibinfo{author}{\bibfnamefont{J.}~\bibnamefont{Binney}},
  \bibinfo{journal}{Mon. Not. Roy. Astron. Soc.}
  \textbf{\bibinfo{volume}{298}}, \bibinfo{pages}{387} (\bibinfo{year}{1998}),
  \eprint{astro-ph/9710077}.

\bibitem[{\citenamefont{Bernabei et~al.}(2008{\natexlab{b}})}]{DAMAApparatus}
\bibinfo{author}{\bibfnamefont{R.}~\bibnamefont{Bernabei}} \bibnamefont{et~al.}
  (\bibinfo{collaboration}{DAMA}), \bibinfo{journal}{Nucl. Instrum. Meth.}
  \textbf{\bibinfo{volume}{A592}}, \bibinfo{pages}{297}
  (\bibinfo{year}{2008}{\natexlab{b}}), \eprint{0804.2738}.

\bibitem[{\citenamefont{Bernabei et~al.}(2008{\natexlab{c}})}]{Bernabei:2007hw}
\bibinfo{author}{\bibfnamefont{R.}~\bibnamefont{Bernabei}}
  \bibnamefont{et~al.}, \bibinfo{journal}{Eur. Phys. J.}
  \textbf{\bibinfo{volume}{C53}}, \bibinfo{pages}{205}
  (\bibinfo{year}{2008}{\natexlab{c}}), \eprint{0710.0288}.

\bibitem[{\citenamefont{Yellin}(2002)}]{Yellin:2002xd}
\bibinfo{author}{\bibfnamefont{S.}~\bibnamefont{Yellin}},
  \bibinfo{journal}{Phys. Rev.} \textbf{\bibinfo{volume}{D66}},
  \bibinfo{pages}{032005} (\bibinfo{year}{2002}), \eprint{physics/0203002}.

\bibitem[{\citenamefont{Gelmini and Gondolo}(2001)}]{Gelmini:2000dm}
\bibinfo{author}{\bibfnamefont{G.}~\bibnamefont{Gelmini}} \bibnamefont{and}
  \bibinfo{author}{\bibfnamefont{P.}~\bibnamefont{Gondolo}},
  \bibinfo{journal}{Phys. Rev.} \textbf{\bibinfo{volume}{D64}},
  \bibinfo{pages}{023504} (\bibinfo{year}{2001}), \eprint{hep-ph/0012315}.

\bibitem[{\citenamefont{Savage et~al.}(2006)\citenamefont{Savage, Freese, and
  Gondolo}}]{Stream}
\bibinfo{author}{\bibfnamefont{C.}~\bibnamefont{Savage}},
  \bibinfo{author}{\bibfnamefont{K.}~\bibnamefont{Freese}}, \bibnamefont{and}
  \bibinfo{author}{\bibfnamefont{P.}~\bibnamefont{Gondolo}},
  \bibinfo{journal}{Phys. Rev.} \textbf{\bibinfo{volume}{D74}},
  \bibinfo{pages}{043531} (\bibinfo{year}{2006}), \eprint{astro-ph/0607121}.

\bibitem[{\citenamefont{Mahapatra}()}]{mahapatra}
\bibinfo{author}{\bibfnamefont{R.}~\bibnamefont{Mahapatra}}, \eprint{Talk at
  PPC 2008}.

\bibitem[{\citenamefont{Aalseth et~al.}(2008)}]{Aalseth:2008rx}
\bibinfo{author}{\bibfnamefont{C.~E.} \bibnamefont{Aalseth}}
  \bibnamefont{et~al.} (\bibinfo{year}{2008}), \eprint{0807.0879}.

\bibitem[{\citenamefont{Tucker-Smith and Weiner}(2001)}]{IDM}
\bibinfo{author}{\bibfnamefont{D.}~\bibnamefont{Tucker-Smith}}
  \bibnamefont{and} \bibinfo{author}{\bibfnamefont{N.}~\bibnamefont{Weiner}},
  \bibinfo{journal}{Phys. Rev.} \textbf{\bibinfo{volume}{D64}},
  \bibinfo{pages}{043502} (\bibinfo{year}{2001}), \eprint{hep-ph/0101138}.

\bibitem[{\citenamefont{Chang et~al.}(2008)\citenamefont{Chang, Kribs,
  Tucker-Smith, and Weiner}}]{IDMNew}
\bibinfo{author}{\bibfnamefont{S.}~\bibnamefont{Chang}},
  \bibinfo{author}{\bibfnamefont{G.~D.} \bibnamefont{Kribs}},
  \bibinfo{author}{\bibfnamefont{D.}~\bibnamefont{Tucker-Smith}},
  \bibnamefont{and} \bibinfo{author}{\bibfnamefont{N.}~\bibnamefont{Weiner}}
  (\bibinfo{year}{2008}), \eprint{0807.2250}.

\bibitem[{\citenamefont{Yao et~al.}(2006)}]{pdg}
\bibinfo{author}{\bibfnamefont{W.~M.} \bibnamefont{Yao}} \bibnamefont{et~al.}
  (\bibinfo{collaboration}{Particle Data Group}), \bibinfo{journal}{J. Phys.}
  \textbf{\bibinfo{volume}{G33}}, \bibinfo{pages}{1} (\bibinfo{year}{2006}).

\end{thebibliography}

\end{document}